\documentclass[prb,twocolumn,superscriptaddress,showpacs,amsmath,amssymb,amsthm]{revtex4}
\usepackage{amsmath,amssymb,color,graphicx,fancyhdr,hyperref,amsthm,units}

\usepackage{graphicx}
\usepackage{latexsym}
\usepackage{amsmath}
\usepackage{amssymb}
\usepackage{amsfonts}
\usepackage{color}
\usepackage{bm}% bold math
\usepackage{verbatim}
\usepackage{epstopdf}

\usepackage{changes}

\usepackage{tikz}
\usepgflibrary{arrows}% for more options on arrows

\begin{document}

\title{Vortex motion and flux-flow resistivity in dirty multiband
superconductors}

\author{Mihail Silaev}
\affiliation{Department of Theoretical Physics, The Royal Institute of
Technology, Stockholm, SE-10691 Sweden}
\author{Artjom Vargunin}
%\email[]{artjom.vargunin}
%\thanks{}
\affiliation{Department of Theoretical Physics, The Royal Institute of
Technology, Stockholm, SE-10691 Sweden}
\affiliation{Institute of Physics, University of Tartu, Tartu, EE-50411,
Estonia}

\date{\today}

\begin{abstract} 
 The conductivity of vortex lattices in multiband superconductors with high
concentration of impurities is calculated based on 
 microscopic kinetic theory. Both the limits of high and low fields  are
considered, when the magnetic induction is close to or much smaller than the
critical field strength $H_{c2}$, respectively. It is shown that in contrast to
 single-band superconductors the resistive properties are not
universal but depend on the pairing constants and ratios of
diffusivities in different bands. The low-field magneto-resistance can strongly 
exceed Bardeen-Stephen estimation in a quantitative agreement with experimental data 
for two-band superconductor MgB$_2$.
 
\end{abstract}

 \maketitle
 \section{Introduction}
 Recent transport experiments reveal quite unusual behaviour of the flux-flow resistive states in multiband superconductors 
 \cite{FluxFlowExperimentsFeAs1, FluxFlowExperimentsFeAs2, FluxFlowExperimentsFeAs3, 
 FluxFlowExperimentsFeAs4,FluxFlowExperimentsFeAs5, FluxFlowExperimentsMgB2}. The magnetic field dependencies of flux-flow resistivity 
 $\rho_f(B)$ were found to be qualitatively different from that observed in single-band superconductors\cite{ExperimentKim}. 
 This behaviour is not explained by the theories developed in previous works. 
 \cite{BS,GorkovKopnin1,GorkovKopnin2,MakiCaroli,Thompson,Ebisawa,Schmid} 

 Vortex motion in conventional type-II superconductors have been investigated for several decades.
 Flux-flow experiments in single-band superconductors at low temperatures and magnetic fields\cite{ExperimentKim}
 are well described by the Bardeen-Stephen (BS) theory\cite{BS}.
  In this regime the flux-flow resistivity is given by the 
 linear magnetic field dependence
 \begin{equation} \label{Eq:BetaDefinition}
 \rho_f/\rho_n= \beta^{-1} B/H_{c2},
 \end{equation}
 where $ \rho_n$ is the normal state resistivity, $B$ is an average magnetic induction, 
 $H_{c2}$ is the second critical field and $\beta\approx 1$.
 The BS law is in the qualitative agreement with the results obtained based on the microscopic 
 theory for dirty superconductors\cite{GorkovKopnin1}. At larger magnetic fields the dependence $\rho_f (B)$ becomes essentially  non-linear\cite{ExperimentKim,Fogel}. At elevated temperatures the significant growth of $\beta$ results in the suppressed  magneto-resistivity as compared to the BS value \cite{GorkovKopnin2,OvchinnikovJETP}.
 
 In strong  magnetic fields $H_{c2}-B \ll H_{c2}$ 
 the motion of dense vortex lattices has been extensively studied 
 with the help of linear response theory \cite{MakiCaroli,Thompson,Ebisawa}. 
 In these works the slope of flux-flow resistivity 
 \begin{equation} \label{Eq:SDefinition}
 S= (H_{c2}/\rho_n) (d\rho_f/dB)
 \end{equation}
 has been shown to have a universal temperature dependence in the dirty limit. 
 It is characterized by a monotonic increase from $S(T=0)=1.72$ to $S(T=T_c)=5$ \cite{Thompson}. 
 This behaviour was confirmed by accurate high-field measurements in Pb-In alloys\cite{ExperimentThompson}.
 The deviation at elevated temperatures near $T_c$ were explained by depairing due to the spin-flip and electron-phonon 
 scattering\cite{LarkinOvchinnikov}.  
       
 In contrast to the conventional behaviour described above, many multiband superconductors\cite{FluxFlowExperimentsFeAs1,FluxFlowExperimentsFeAs2,  FluxFlowExperimentsFeAs3, FluxFlowExperimentsFeAs4, 
 FluxFlowExperimentsFeAs5}  including MgB$_2$
 \cite{FluxFlowExperimentsMgB2} 
 were found to have the flux-flow resistivity larger than the BS value 
 $\rho_{f}/\rho_n > B/H_{c2}$. The experimentally found dependencies $\rho_f(B)$
have a steeper growth in the low-field region with an enhanced magneto-resistance characterized by $\beta< 1$ 
and a smaller slope $S<1$ at
$B=H_{c2}$\cite{FluxFlowExperimentsMgB2}, which is not described by the
single-band theory\cite{Thompson}. 
  
 The existing theories of flux-flow states cannot be straightforwardly applied 
 to multiband superconductors. In these systems vortices have a composite 
 structure consisting of multiple singularities corresponding to the order parameter 
 phase windings in different superconducting bands. In equilibrium an isolated 
 composite vortex is a bound state of several co-centred fractional 
 vortices \cite{babaev2}. They can split however under the action of 
 fluctuations\cite{babaev3}, interaction with other vortices and sample 
 boundaries \cite{dao,silaev} or due to external drive\cite{lin}.
 In particular it was shown that the moving composite vortices can split into 
 separate fractional vortices and even dissociate in a 
 non-linear regime provided the interband pairing is sufficiently small\cite{lin}. 
 It is natural to expect that vortex splitting should have a profound effect on the 
 flux-flow resistivity especially at high fields when the flux-flow resistivity is strongly affected by
the distortions of the moving vortex lattice\cite{MakiCaroli,Thompson,Ebisawa}. As will be shown below,
the well-known solution\cite{Schmid} describing moving vortex lattice  
is not applicable to describe multiband systems since the distortion generically splits the sublattices of fractional vortices. In the present paper we develop a theoretical framework
to take into account this effect and calculate the conductivity corrections. For that one needs to know the Maki parameter also known as a generalized Ginzbirg-Landau parameter $\kappa_2$ which determines in particular the order parameter density as a function of magnetic field near $H_{c2}$\cite{MakiCaroli}. Recently this parameter has been calculated for multiband superconductors\cite{SilaevKappa2}.  
  
 To obtain a complete picture of the flux-flow conductivity behaviour in 
 multiband systems we consider also the regime of small magnetic fields, 
 when a picture of isolated moving vortices 
 is an adequate description\cite{GorkovKopnin1}. Based on the kinetic theory we 
 calculate the  
 coefficient $\beta$ which characterizes the initial slope of the magnetoresistance.
 Applying the combination of the results in two limiting cases of small and high 
 magnetic fields it is possible to fit the experimental curves $\rho_f(B)$ 
 for multiband superconductors with known pairing interactions such as MgB$_2$.
  
 The model of dirty-limit superconductors assumed in the present work is appropriate 
 for a certain class of multiband materials including MgB$_2$\cite{GurevichMgB2,KoshelevGolubovHc2} and iron-pnictides\cite{GurevichFeAsNature}. 
 In single crystals of MgB$_2$ the de Haas-van Alphen data \cite{dHvAMgB2} and 
 thermal conductivity 
 measurements \cite{ThCondMgB2} suggest the borderline regime when one 
 of the superconducting bands is moderately clean and the other one is moderately 
 dirty\cite{KoshelevGolubovHc2}. Scanning tunnel microscopy  
 shows absence of zero-bias anomaly inside vortex cores which is typical 
 for dirty superconductors\cite{STMMgB2}. Impurities at high concentration can 
 be introduced in MgB$_2$ on demand during the preparation process 
 producing non-trivial magnetic properties which have been intensively studied 
 recently \cite{GurevichMgB2,MgB2Ex1,MgB2Ex2,MgB2Ex3,MgB2Ex4}. 
     
 The structure of this paper is as follows. In Sec.\ref{Sec:KinEq} we introduce 
 the  Keldysh-Usadel description of the kinetic processes in dirty 
 multiband superconductors. Here the basic components of the kinetic theory are 
 discussed including kinetic equations, self-consistency equations for the order 
 parameter and current as well as a general expression for the viscous force acting 
 on the moving vortices. The flux-flow conductivity at high magnetic fields
 is calculated in Sec.\ref{Sec:LargeFields} taking into account the splitting of 
 fractional vortex sublattices. The case of low fields is considered in 
 Sec.\ref{Sec:SmallFields}. Quantitative comparison of theoretical results
 with flux-flow resistivity measurements in MgB$_2$ \cite{FluxFlowExperimentsMgB2} 
 is discussed in Sec. \ref{Sec:ExamplesComparison}. 
 The work summary is given in Sec.(\ref{Sec:Conclusion}).
               
 \section{Kinetic equations and forces acting on the moving vortex line}
\label{Sec:KinEq}
    
 We consider multiband superconductors in a dirty limit when the kinetics and
spectral properties are described by the Keldysh-Usadel theory. 
 For the single band case the theory of vortex motion in diffusive
superconductors was developed in works
\cite{GorkovKopnin1,GorkovKopnin2,LarkinOvchinnikov,Thompson}. Here we
generalize their theory to the multiband case.  
 
 The quasiclassical GF in each band is defined as
 \begin{equation}
 \check{g}_k = \left(%
 \begin{array}{cc}
  \hat g^R_k &  \hat g^K_k \\
  0 &  \hat g^A_k \\
 \end{array}\label{eq:GF0}
 \right)\; ,
 \end{equation}
 where $g^K_k$ is the (2$\times$2 matrix) Keldysh component 
 and $\hat g^{R(A)}_k$ is the retarded (advanced) GF, $k$ is the band index. In
dirty superconductors the matrix $\check g$
 obeys the Usadel equation 
 \begin{equation}\label{Eq:KeldyshUsadelText}
 \{\tau_3\partial_t, \check g_k \}_t = D_k\hat\partial_{\bf r}  ( \check g_k
\circ \hat\partial_{\bf r} \check g_k) 
 + [\hat H_k , \check g_k ]_t -i [\check \Sigma^{ph} , \check g_k ]_t  
  \end{equation} 
 where $D_k$ is the diffusion constant, and  
 $\hat H_k ({\bm r},t) = i \hat \Delta_k $ 
 where $\hat\Delta_k (t)= i |\Delta_k| \tau_2 e^{-i\theta_k \tau_3}$ 
 is the gap operator in $k$-th band. We use from the beginning the temporal
gauge where the scalar potential 
 is zero $\Phi=0$ with and additional constraint that
 in equilibrium the vector potential is time-independent and satisfies
$\nabla\cdot\bm A=0$.  
 In Eq.(\ref{Eq:KeldyshUsadelText}) the covariant differential superoperator is
defined by 
  $$
  \hat \partial_{\bf r} \hat g_k= \nabla \hat g_k -ie
  \left[\tau_3{\bm A}, \hat g_k \right]_t .
  $$ 
  
 The gap in each band is determined by self consistency equation
 \begin{equation}\label{Eq:SelfConst}
 \Delta_k (t,\bm r) = \frac{\pi}{2}\lambda_{kj} 
 (\hat g^{K}_j)_{12}(t_{1,2}=t,\bm r). 
 \end{equation}   
 Here $\hat\Lambda $ is the coupling matrix satisfying symmetry relations
$\nu_1\lambda_{12}=\nu_2\lambda_{21}$, where $\nu_k$ is the DOS.   
 The electric current density is given by 
 \begin{equation}\label{Eq:Current}
 {\bm j} (t,\bm r) =\frac{\pi e}{2}\sum_k \nu_k D_k {\rm Tr} 
 (\check g_k \circ\hat \partial_{\bm r} \check g_k )^K (t_{1,2}=t,\bm r) 
 \end{equation} 
  We define the commutator operator as
  $[X, g]_t= X(t_1) g(t_1,t_2)- g(t_1,t_2) X(t_2)$, similarly for anticommutator 
  $\{,\}_t$. We introduce the symbolic product operator 
  $ (A\circ B) (t_1,t_2) = \int dt A(t_1,t)B(t,t_2) $.
      
   The Keldysh-Usadel Eq. (\ref{Eq:KeldyshUsadelText}) is complemented by 
   the normalization condition
   $(\check g_k\circ \check g_k) (t_1,t_2)= \check \delta (t_1 - t_2)$
   which allows to introduce parametrization of the Keldysh component in 
   terms of the distribution function
   \begin{eqnarray}
   \hat g^K_k = \hat g^R_k\circ \hat f^{(k)}- \hat f^{(k)}\circ \hat g^A_k \\
   \label{Eq:Parametrization}
   \hat f^{(k)}= f^{(k)}_L\tau_0 +f^{(k)}_T\tau_3 .  \label{Eq:DistrFun}
   \end{eqnarray}
   To proceed we introduce the mixed representation in the time-energy 
   domain as follows
  $g_k(t_1,t_2) = \int_{-\infty}^{\infty} g_k(\varepsilon, t) 
  e^{-i\varepsilon(t_1-t_2) } d\varepsilon/(2\pi) $, 
  where $t=(t_1+t_2)/2$. 
   
 %%%%%%%%%%%%%%%%%%%%%%%%%%%%%%%%%%%%%%%%
 %%%%%%%%%%%%%%%%%%%%%%%%%%%%%%%%%%%%%%%%
 
 We employ Larkin-Ovchinnikov theory\cite{LarkinOvchinnikov,KopninBook} to
calculate the force acting on the moving vortex line. In multiband
superconductors the force is given by a linear superposition of contributions from different bands\cite{KopninBook,LarkinOvchinnikov}
 \begin{eqnarray} \label{Eq:FenvGen}
 {\bm F}_{env}= \sum_k {\bm F}^{(k)}_{env} + \frac{1}{c} \int d^2 {\bm r} 
 {\bm H}\times{\bm j}^{(nst)} \\ \label{Eq:FenvPartial} 
 {\bm F}^{(k)}_{env} = \nu_k \int d^2 {\bm r}  \int_{-\infty}^{\infty}
 \frac{d\varepsilon}{4} 
 {\rm Tr} (\hat g^{nst}_k \hat \partial_{\bm r} \hat \Delta_k) 
 \end{eqnarray} 
 where the covariant differential superoperator is given by 
 $ \hat \partial_{\bf r} \hat \Delta_k= \nabla \hat \Delta_k -ie
\left[\tau_3{\bm A}, \hat \Delta_k \right] $.
 Eqs.(\ref{Eq:FenvGen},\ref{Eq:FenvPartial} ) contain a non-stationary part of
 the electric current density  ${\bm j}^{(nst)}$ and the Keldysh component of 
 a non-stationary Green's function $\hat g^{nst}_k$ which can be expressed
 through the gradient expansion as follows 
 \begin{eqnarray}\label{Eq:GnstExp}
 \hat g^{nst}_k = -\frac{i}{2} \partial_t (\hat g^R_k + \hat
 g^A_k)\partial_\varepsilon f_0 + \\
 (\hat g^R_k - \hat g^A_k )f^{(k)}_L +  (\hat g^R_k\tau_3 - \tau_3\hat
 g^A_k)f^{(k)}_T.
 \end{eqnarray}
  
 Keeping the first order non-equilibrium terms and neglecting the
 electron-phonon relaxation in Eq.(\ref{Eq:KeldyshUsadelText}) we obtain a system 
 of two coupled kinetic equation that determine both the transverse $f_T^{(k)}$ 
 and longitudinal $f_L^{(k)}$ distribution function components
\begin{align} \nonumber
 & \nabla ( {\cal D}_T^{(k)} \nabla f_{T}^{(k)} )+  {\bm j}_e^{(k)}\cdot\nabla f_L^{(k)} + 
 2i {\rm Tr} [ (\hat g^R_k + \hat g^A_k) \hat \Delta_k ] f_T^{(k)} = \\ \label{Eq:KineticEqFT3} 
 & \partial_\varepsilon f_0 {\rm Tr} [ \tau_3  \partial_t\hat \Delta_k( \hat g^R_k +
\hat g^A_k) ] -
 e \partial_\varepsilon f_0 \nabla \cdot ({\cal D}_T^{(k)} {\bm E} ),\\ 
 %\end{eqnarray} 
 %\begin{align}
 \nonumber
 &\nabla ( {\cal D}_L^{(k)} \nabla f_{L}^{(k)} ) + {\bm j}_e^{(k)}\cdot\nabla f_T^{(k)} + 2i {\rm Tr}
 [\tau_3 (\hat g^R_k -\hat g^A_k) \hat \Delta_k ] f_T^{(k)} = \\  \label{Eq:KineticEqFL} 
 & - \partial_\varepsilon f_0 {\rm Tr}[ \partial_t\hat \Delta_k( \hat g^R_k - \hat
 g^A_k) ] - e \partial_\varepsilon f_0  
 {\bm j}_e^{(k)} \cdot {\bm E}, 
 \end{align} 
where ${\bm E}$ is the electric field, the energy-dependent diffusion
 coefficients ${\cal D}_{T,L}^{(k)}$
 and the spectral charge current ${\bm j}_e^{(k)}$ in each band are given by 
 \begin{eqnarray} \label{Eq:DiffCoeffT} 
 {\cal D}^{(k)}_T = D_k {\rm Tr}(\tau_0 - \tau_3\hat g^R_k \tau_3\hat g^A_k ) \\
 \label{Eq:DiffCoeffL}
 {\cal D}^{(k)}_L = D_k {\rm Tr} (\tau_0 - \hat g^R_k \hat g^A_k ) \\
 \label{Eq:SpectralCurrent}
 {\bm j}_e^{(k)} = D_k {\rm Tr}\; [ \tau_3( \hat g^R_k\hat\partial_{\bf r} 
 \hat g^R_k - \hat g^A_k\hat\partial_{\bf r} \hat g^A_k)]   
 \end{eqnarray}
 The detailed derivation of this system is given in the Appendix\ref{App:KinDerivation}. 
 In the simplest case of weak spatial inhomogeneities it coincides with the equations 
 derived by Sch\"{o}n \cite{Schon} with the substitution $f^{(k)}_T \rightarrow
 f^{(k)}_T + (\dot{\theta}_k/2-e\Phi)\partial_\varepsilon f_0 $.   
             
 \section{Large magnetic fields. }  \label{Sec:LargeFields}
 At large magnetic fields $H_{c2}- H \ll H_{c2}$ we can use simplifying
 approximations related to the smallness of the order parameter  
 $|\Delta_k|\propto \sqrt{ 1- H/H_{c2}}$. From the kinetic
 Eqs.(\ref{Eq:KineticEqFT3} ) one can see that 
 non-equilibrium distribution functions are by the order of magnitude
 $f^{(k)}_{L,T} \propto |\Delta_k|^2$. 
 Therefore the contribution to the force  determined by $g^{nst}_k$ is
 proportional to $|\Delta_k|^3$. 
 Hence the force on the moving vortex is determined by the second term in
 Eq.(\ref{Eq:FenvGen}) containing a  non-stationary 
 part of the current ${\bm j}^{nst}$ which is proportional to $|\Delta_k|^2$ as
 will be shown below. 
 The most efficient way to find ${\bm j}^{nst}$ 
 is to calculate the total current and then extract a non-stationary part
 proportional to the vortex velocity ${\bm v}_L$.
 From  Eq.(\ref{Eq:Current}) we have 
 \begin{equation}\label{Eq:CurrentHc2}
 {\bm j} =  \sum_k \frac{e\nu_k}{4} \int_{-\infty}^{\infty} d\varepsilon 
 \left( {\bm j}_e^{(k)} f_0  + e {\bm E} {\cal D}^{(k)}_T 
 \frac{\partial f_0}{\partial \varepsilon}  \right).
 \end{equation}    
 The force balance condition yields that the space-averaged net 
 current (\ref{Eq:Current}) is equal to the external transport current
 $\langle{\bm j}\rangle={\bm j}_{tr}$.  
 In equilibrium superconducting currents circulate around stationary vortices so
 that the net current is zero. 
 Under the non-equilibrium conditions created by moving vortices both the two
 terms in the Eq.(\ref{Eq:CurrentHc2}) 
 provide non-zero contributions.
 Hereafter we will consider isotropic superconductors so that the average
 current is co-directed with the electric field 
 $\langle {\bm j}\rangle = \sigma_f {\bm E}$.  The flux flow conductivity
 is given by the superposition of three terms
 $\sigma_f=\sigma_n+\sigma_{fl}+\sigma_{st}$, where 
 $\sigma_n$  is a large normal-metal contribution and the last two terms are
 given by       
 \begin{align} \label{Eq:CurrentSupercondDistortions}
 \sigma_{fl} =  \sum_k \frac{e\nu_k}{4E} \int_{-\infty}^{\infty} \langle {\bm
 j}_e^{(k)}\rangle f_0(\varepsilon)d\varepsilon   \\ 
 \label{Eq:CurrentQuasiparticle}
 \sigma_{st} =  - \sum_k \frac{e^2\nu_k}{4} \int_{-\infty}^{\infty} 
 \frac{\partial\langle {\cal D}^{(k)}_T \rangle}{\partial\varepsilon}
f_0(\varepsilon) d\varepsilon 
 \end{align}   
 The term $\sigma_{fl}$ (\ref{Eq:CurrentSupercondDistortions}) is a conductivity
 correction generated by 
 non-equilibrium distortions or fluctuations of the superconducting order parameter. 
 The similar correction in single-band superconductors has been calculated in
 the pioneering works on the 
 flux-flow conductivity at $H\approx H_{c2}(T)$\cite{MakiCaroli,Thompson,Ebisawa}.
 Besides  there exists a sizeable quasiparticle contribution
 to the current given by the second term in Eq.(\ref{Eq:CurrentHc2})
 which determines the conductivity correction $\sigma_{st}$\cite{Thompson,Ebisawa}. 
 As can be seen from the
 Eq.(\ref{Eq:CurrentQuasiparticle}) this correction is generated by  
 nonequilibrium quasiparticles and the
 superconductivity-induced changes of the 
 diffusion coefficient ${\cal D}^{(k)}_T-4D_k$ as compared to the normal state
 which has ${\cal D}^{(k)}_T=4D_k$.
 In contrast to $\sigma_{fl}$ the quasiparticle contribution $\sigma_{st}$ can
 be calculated using the static order parameter distribution.  
       
 \subsection{ Conductivity correction $\sigma_{fl}$. }       
 To calculate $\sigma_{fl}$ given by (\ref{Eq:CurrentSupercondDistortions}) 
 we need to find corrections to the order parameter fields 
 $\Delta_k$ in a moving Abrikosov lattice. In single-band superconductors such 
 corrections were calculated in works \cite{MakiCaroli,Ebisawa,Schmid}. 
 The analogous problem in multiband superconductors cannot be approached using 
 a straightforward generalization of the single-band solution due to the complex 
 structure of vortices in multiband superconductors which are composite objects 
 consisting of several overlapping fractional vortices in different bands. 
      
 %%%%%%%%%%%%%%%%%%%%%%%%%%%%%%%%%%%%%%%%%%%%%%%%%%%%
 \begin{figure*}[htb!] \includegraphics[width=0.8\linewidth]{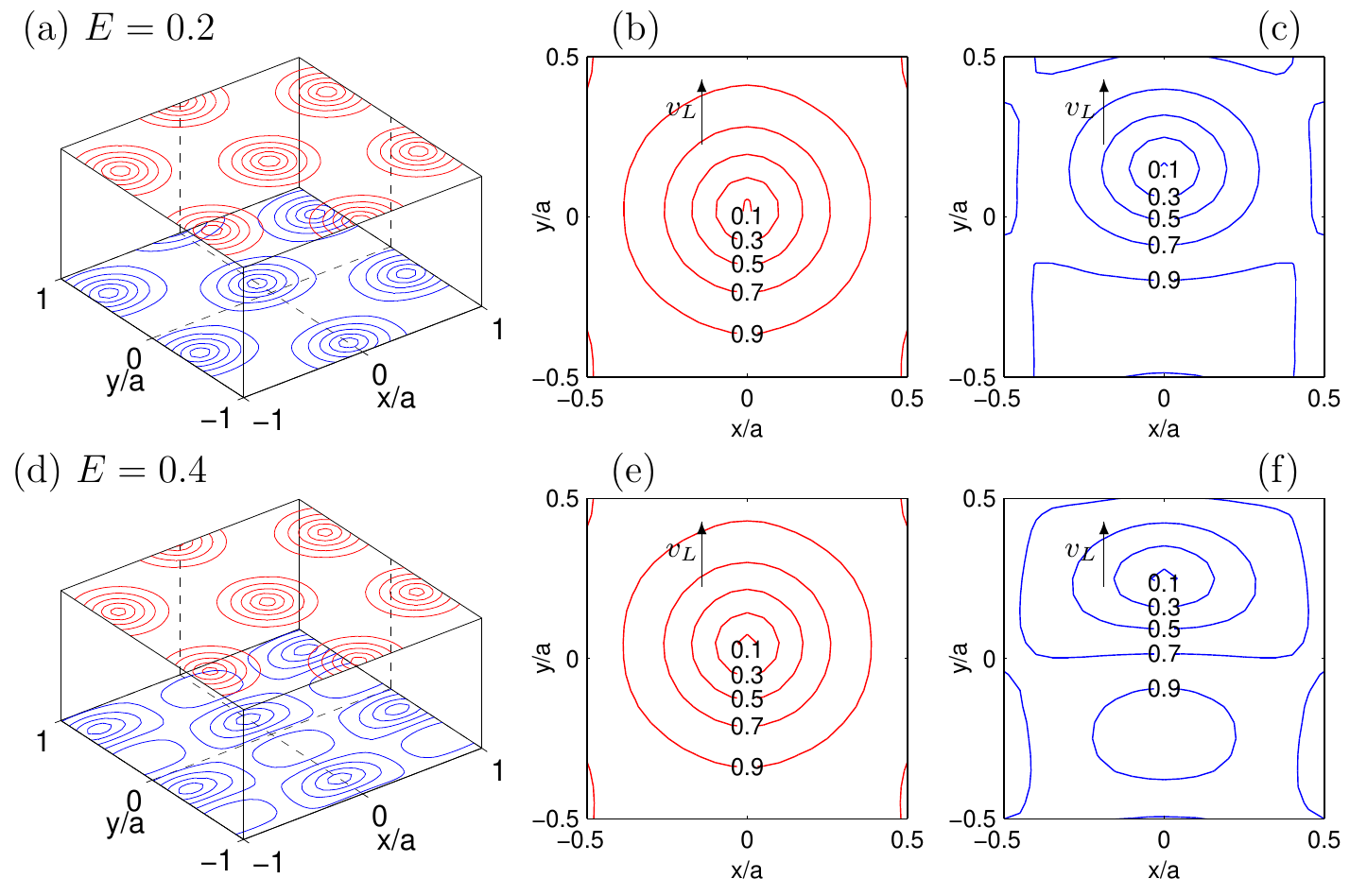} 
 \caption{\label{Fig:VortLattice} (Color online) 
 Distorted lattices of fractional vortices moving 
 along the $y$ direction in a two-band superconductor MgB$_2$ with $D_2=0.05D_1$
and 
 the pairing constants mentioned in the text. 
 The temperature is $T=0.1 T_c$. % and $e=0.05$.
 The transport current is applied along $x$. 
 The magnitude of electric field is normalized by $T_c/(e\xi_1)$, where 
 $\xi_1=\sqrt{D_1/T_c}$. The length is normalized to the triangular lattice
spacing 
 $a= 2L_H(\pi/\sqrt{3})^{1/2}$.
 (a,d) Density levels of $\Delta_1$ ($\Delta_2$) in upper (lower) plots for
$E=0.2, \; 0.4$. 
  Detailed density levels of $\Delta_1$, $\Delta_2$ are shown in (b,d) for
$E=0.2$ and
 in (e,f) for $E=0.4$.  The vortex velocity direction $\bm {v}_L$ is shown by
arrows in (b,c,e,f).
  }
 \end{figure*}
 %%%%%%%%%%%%%%%%%%%%%%%%%%%%%%%%%%%%%%%%%%%%%%%%%%%%       
    
 To calculate the structure of moving vortex lattice {in} a two-band
superconductor 
 let us consider the 
 linear integral-differential system of linearised non-stationary Usadel
 equations together with self-consistency Eq.(\ref{Eq:SelfConst}) for the order
parameter  
 \begin{align} \label{Eq:UsadelLinearized} 
 & \frac{D_k}{2}\left( \nabla - 2ie {\bm A} \right)^2 f^{R,A}_k \pm i\varepsilon
f^{R,A}_k = i\Delta_k, \\ \label{Eq:SelfConstLinearized}  
 %\frac{D_k}{2}{\bm \Pi}^2 f^{A}_k - i\varepsilon f^{A}_k = i\Delta_k  \\
 & \Delta_k =\sum_j\frac{\lambda_{kj}}{4} \int\limits_{-\infty}^{\infty} d\varepsilon  \left[
f^R_j- f^A_j +
 \frac{i}{2} \frac{\partial^2}{\partial t\partial \varepsilon} (f^R_j +
f^A_j)\right] f_0(\varepsilon).
 \end{align}   
To derive the self-consistency Eq. (\ref{Eq:SelfConstLinearized}) we substituted
the Keldysh component expansion
$\hat g^K_k=(\hat g^R_k-\hat g^A_k)f_0 - \frac{i}{2}(\hat g^R_k+\hat
g^A_k)\partial_\varepsilon f_0$ {and} integrated the second term by parts. 
The vector potential in Eq. (\ref{Eq:UsadelLinearized}) describes a uniform
magnetic field ${\bm B}= H_{c2} {\bm z}$ 
and a uniform electric field in ${\bm x}$ direction so that ${\bm A}= {\bm y}
H_{c2} x - {\bm x} E t $.
It is more convenient for calculations to remove a non-stationary part 
 of the vector  potential by a gauge transform introducing scalar potential
$\Phi = - Ex$. Then the time derivative 
  in Eq.(\ref{Eq:SelfConstLinearized}) elongates $\partial_t\rightarrow
\partial_t + 2ie \Phi $.  
  
  A periodic vortex lattice moving with the  constant velocity ${\bm v}_L=
v_L{\bm y}$ is described by the following 
  solution of Eqs.(\ref{Eq:UsadelLinearized},\ref{Eq:SelfConstLinearized} ) 
  \begin{eqnarray}\label{Eq:Ansatz}
  \Delta_k= \sum C_n e^{ i n p(y-v_Lt) } \tilde{\Delta}_k(x-nx_0), \\
  f^{R,A}_k = \sum C_n e^{ i n p(y-v_Lt)} \tilde{f}^{R,A}_k(x-nx_0),
  \end{eqnarray}
   where $|C_n|=1$,  $x_0= p/(2eH_{c2})$ and the parameter $p$ is determined by
the lattice geometry. 
   The vortex velocity should satisfy $v_L= - E/H_{c2}$ in order for the
solution to keep magnetic translation symmetry in $x$ direction.
   Substituting ansatz (\ref{Eq:Ansatz}) into Eq.(\ref{Eq:UsadelLinearized}) we
get  
   \begin{equation}
   \frac{D_k}{2} \hat L_x f^{R,A}_k \pm  i\varepsilon f^{R,A}_k = i\Delta_k,  
    \end{equation}
    where $\hat L_x = \partial^2_x -(2eH_{c2})^2 x^2 $.
    One can see that the principal difference with a single component is due to the different 
    diffusion constants which do not allow the solution to have a form of shifted zero Landau level eigenfunction. 
    Instead we should search it as a superposition of 
  \begin{eqnarray}
  \tilde{f}^{R,A}_k &=& a^{R,A}_{k0} \Psi_0 (x)+ a^{R,A}_{k1} \Psi_1 (x) \\
  \tilde{\Delta}_k &=& b_{k0}\Psi_0 (x)+ b_{k1}\Psi_1 (x)
  \end{eqnarray}       
  where $\Psi_0(x) = \exp(-x^2/2L_H^2)$ and $ \Psi_1 (x) = x \Psi_0(x) $
  satisfy $\hat L_x\Psi_0 = - \Psi_0/L_H^{2} $ and {$\hat L_x\Psi_1 =
-3\Psi_1/L_H^{2}$}. 
  Since the admixture of the first LL is proportional to a small parameter
$E/H_{c2}$ 
  we can determine the coefficients $a_0$, $b_0$ using a stationary equation
Eq.(\ref{Eq:UsadelLinearized})
  \begin{eqnarray} \label{Eq:Coeffitient0}
   a^{R,A}_{k0} &=&  b_{k0}/(iq_k \pm \varepsilon), \\ \label{Eq:Coeffitient1}
   a^{R,A}_{k1} &=& b_{k1}/(3iq_k \pm \varepsilon) ,
   \end{eqnarray}
   where $q_k = e H_{c2} D_k$. 
   Substituting the relation (\ref{Eq:Coeffitient0}) to the self-consistency
Eq.(\ref{Eq:SelfConstLinearized})
   yields a homogeneous linear equation    
  \begin{align} \label{Eq:Hc2}
   &\hat A {\bm b}_0 =0,  \\\label{Eq:Hc2_A}
   \hat A = \hat\Lambda^{-1} -& {\tau_0\left[ G_0 - \ln(t) + \psi (1/2)\right]+\psi (1/2+\hat\rho)}, 
   \end{align}
   where $G_0$ is the minimal positive eigenvalue of the inverse coupling matrix, 
   $t=T/T_c$, $(\hat\rho)_{ik}= \delta_{ik} \rho_k$ and $\rho_k=q_k/2\pi T$.
   The solvability condition ${\rm det} \hat A =0$ determines the second
critical field of a multiband superconductor. The amplitudes $b_{k1}$ of the first LL admixture are determined substituting
Eq.(\ref{Eq:Coeffitient1}) into the 
    self-consistency Eq.(\ref{Eq:SelfConstLinearized})
    \begin{align} \label{Eq:VortexShift}   
    {\bm b}_1 = \frac{ieE}{2\pi T} \hat A_1^{-1} \psi^\prime(1/2+\hat\rho)
\bm{b}_{0}, \\ \nonumber
     \hat A_1 = \hat A + \psi (1/2+3\hat\rho) - \psi (1/2+\hat\rho).
    \end{align} 
The above relation between components of vectors ${\bm b_1}$ and ${\bm b_0}$ characterizes splitting of composite vortex into separate constituents. Generally, splitting is present for any non-degenerate multiband superconductor having different diffusivities and 
coupling constants in the bands. By increasing strength of electric field, distortion of vortex lattice becomes more evident, see Fig. (\ref{Fig:VortLattice}).
           
 The moving lattice distortions is induced by the first LL admixture in
Eq.(\ref{Eq:VortexShift}) 
 which generates a finite net current perpendicular to the vortex velocity. 
 From Eq.(\ref{Eq:CurrentSupercondDistortions}) we obtain  
  \begin{align} \label{Eq:CurrentFl}
 &\sigma_{fl} = \sum_k  \frac{\langle |\Delta_k|^2 \rangle}{4\pi Te E}
\frac{\sigma_k}{\rho_k} 
 \frac{{\rm Im} (b^*_{k0}b_{k1})}{|b_{k0}|^2} \left[\psi(1/2+3\rho_k)-\psi_k\right], 
  \end{align} 
  where $\psi_k =\psi(1/2+\rho_k)$ and the average order parameter amplitude is given by 
 $\langle |\Delta_k|^2 \rangle = \sqrt{\pi} |b_{k0}|^2L_H/x_0$. 
      
 \subsection{ Conductivity correction $\sigma_{st}$. }      
       
 To calculate the second term contribution in (\ref{Eq:Current}) giving the
conductivity correction 
 $\sigma_{st}$ (\ref{Eq:CurrentQuasiparticle}) we need to find out how the
diffusion coefficients $ {\cal D}^{(k)}_T$ are modulated by the 
 vortex lattice. For this we determine spectral functions $\hat g^{R,A}_k$ 
  using the linearised Usadel Eq. (\ref{Eq:UsadelLinearized}) supplemented by
the normalization condition $(\hat g^{R,A}_k)^2 =1$:
  \begin{align} \label{Eq:SpectralFunctions}
  &\hat g^R_k = \left[ 1 + \frac{|\Delta_k|^2}{2(iq_k+\varepsilon)^2}\right]
\tau_3 + 
   \frac{i|\Delta_k|\tau_2e^{-i\varphi_k\tau_3 }}{iq_k+\varepsilon} \\
  &\hat g^A_k = -\left[ 1 + \frac{|\Delta_k|^2}{2(iq_k-\varepsilon)^2}\right]
\tau_3 + 
   \frac{i|\Delta_k|\tau_2e^{-i\varphi_k\tau_3 }}{iq_k-\varepsilon}  . 
  \end{align}
    Substituting these expressions to the Eq.(\ref{Eq:DiffCoeffT} ) we get
  \begin{equation}\label{Eq:DiffCoeff}
  {\cal D}^{(k)}_T = 2D_k \left[ 1 + \frac{|\Delta_k|^2}{2q_k(q_k+i\varepsilon )} - 
  \frac{|\Delta_k|^2}{2(q_k+i\varepsilon )^2} +c.c. \right] 
  \end{equation}
       
   Using Eq.(\ref{Eq:DiffCoeff}) we evaluate the conductivity  correction
(\ref{Eq:CurrentQuasiparticle}) as follows
   \begin{equation}\label{Eq:CurrentElField}
   \sigma_{st} = \sum_k \frac{\sigma_k\langle|\Delta_k|^2\rangle}{8\pi^2 T^2 }
   \left( \frac{\psi^{\prime}_k}{\rho_k} + \psi^{\prime\prime}_k  \right),   
   \end{equation}
   where %$\rho_k = eH_{c2}D_k/(2\pi T)$ and 
   $\psi^{(n)}_k=\psi^{(n)}(1/2+\rho_k)$ and the partial conductivities are 
   $\sigma_k = 2e^2\nu_kD_k$. One can see that the quasiparticle current 
   (\ref{Eq:CurrentElField}) is given by the superposition of two 
   single-band contributions.
  
 \subsection{Slope of the flux-flow resistivity at $H=H_{c2}(T)$.}  
   \label{Sec:SHc2}  
 We have found that both the conductivity corrections $\sigma_{fl}$ and $\sigma_{st}$ (\ref{Eq:CurrentFl},\ref{Eq:CurrentElField}) 
 are proportional to the average order parameter $\langle|\Delta_k|^2\rangle$ which should be expressed through the magnetic
 field. The average gap functions $\langle|\Delta_k|^2\rangle=\Delta^2 a^2_k$ have a common amplitude which have been calculated 
 in the Ref.(\cite{SilaevKappa2}) 
 \begin{equation} \label{Eq:OrderParameterAmplitude}
 \Delta= \left( \frac{eT\delta H}{2\beta_L} 
 \frac{\sum_{k} \nu_k a^2_k D_k\psi^{\prime}_k }
 {\sum_{k} \nu_k a^4_k \sigma_kD_k\psi^{\prime 2}_k \tilde{\kappa}_k^2 } \right)^{1/2},
 \end{equation}
 where $\delta H = H_{c2}-H$ and $\beta_L$ is an Abrikosov parameter equal to $1.16$ for a triangular lattice\cite{AbrikosovParameter}. 
 The parameters $\tilde{\kappa}_{k}$ which in the single band case characterize
 the magnetization slope at $H_{c2}(T)$ \cite{Maki,Caroli} are given by   
 \begin{equation}\label{Eq:kappaSingleBand}
 \tilde{\kappa}_{k} = \left( \frac{-\psi^{\prime\prime}_k}
 {16\pi\sigma_kD_k\psi^{\prime 2}_k} \right)^{1/2}.
 \end{equation}   
 The coefficients $a_k$ are determined unambiguously by the Eq.(\ref{Eq:Hc2})
 supplemented by a normalization condition $\sum_k a_k^2 =1$ so that 
 %$a_k = |b_{k0}|/\sqrt{\bm b_{0}^2}$. 
$a_k = |b_{k0}|/\sqrt{\sum_k |b_{k0}|^2}$.
    
\begin{figure}[h!]\includegraphics[width=1.0\linewidth]{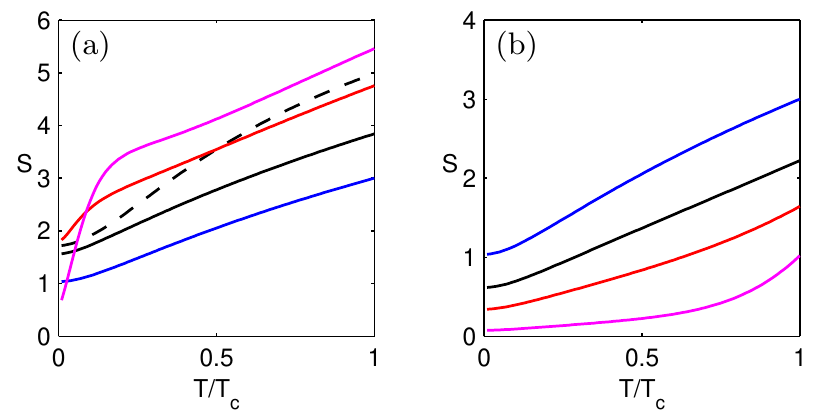} 
  \caption{\label{Fig:Conductivity} (Color online) 
 Slope of the flux-flow resistivity $S=-(H/\sigma_n)d\sigma_f/dH$ at
 $H=H_{c2}(T)$ as a function of temperature $T$ 
 for different values 
 of the ratio between band diffusivities $\eta=D_1/D_2$. (a) Solid curves from
 top to bottom $\eta=20;\;5;\;2;\;1 $. 
 The dashed line shows a universal single-band behaviour\cite{Thompson,Ebisawa}.
 (b) Curves from top to bottom $\eta=1;\;0.5;\; 0.25;\;0.05$. 
 The pairing parameters are $\lambda_{11}=0.101; \; \lambda_{22}=0.045;\; 
 \lambda_{12}=0.034;\; \lambda_{21}=0.026$ \cite{KoganMgB2}.  }
 \end{figure}

 Substituting the order parameter amplitude (\ref{Eq:OrderParameterAmplitude})
 to the expressions for conductivity corrections 
 (\ref{Eq:CurrentSupercondDistortions},\ref{Eq:CurrentQuasiparticle})
 we can find the flux-flow conductivity slope at $H=H_{c2}(T)$ (\ref{Eq:SDefinition}) 
 which can by written in the form {$S = -(H_{c2}/\sigma_n) d\sigma_f/dH$}. 
 In contrast to the dirty single-band superconductors which are characterized
 by a universal $S=S(T)$ curve the multiband superconductors have a significant
 variation of $S$ as a  function of the ratio between band diffusivities $D_1/D_2$. 
 The sequence of $S(T)$ dependencies for different values of $D_1/D_2$ are
 shown in Fig. (\ref{Fig:Conductivity}) for the two-band superconductor with pairing constants corresponding to the weak coupling model of
 MgB$_2$\cite{KoganMgB2}. 
 For the reference the universal single-band curve is shown by the dashed line
 in Fig.(\ref{Fig:Conductivity})a.        
 
 %%%%%%%%%%%%%%%%%%%%%%%%%%%%%%%%%%%%%%%%%%%%%%%%%%%%
 %%%%%%%%%%%%%%%%%%%%%%%%%%%%%%%%%%%%%%%%%%%%%%%%%%%%
 
By applying our model at elevated temperatures, we neglect interband impurity scattering assuming that it is much smaller compared to the orbital depairing energy 
   $eD_kH_{c2}$. Due to the same reason we omit scattering at paramagnetic impurities and inelastic electron-phonon relaxation\cite{LarkinOvchinnikov} which are known to be important near $T_c$ but are negligible at lower temperatures.

 \subsubsection{ Limiting values of $S$ at temperatures close to $T_c$ }
   
 Qualitatively the significant deviations of $S(T)$ dependencies from the 
 single-band  case can be understood analysing limiting case of $T\to T_c$
 when the critical field is small so that $\rho_k\to 0$ and one can use the 
 asymptotic values of functions $\psi^\prime_k= \pi^2/2$, 
 $\psi^{\prime\prime}_k =- 14\zeta(3)$. In this case
 the splitting of fractional vortex sub-lattices vanishes. 
 As can be seen from  Eqs.(\ref{Eq:VortexShift}) to the first order by $\rho_k$ we have   
 \begin{equation}
 {\bm b_1}= \frac{ieE}{4\pi T}
 \frac{ {\rm Tr} \hat A }{ \sum_k A_{kk}\rho_k } {\bm b_0}.
 \end{equation}
 This expression means that current-driven fractional vortices in different
 bands shift by the same amount.
    
  The conductivity corrections are given by
  \begin{align}
  &\sigma_{fl}=\frac{{\rm Tr}\hat A }{ \sum_k A_{kk}\rho_k } 
  \sum_k \frac{\sigma_k\langle|\Delta_k|^2\rangle}{16 T^2 } \\
  &\sigma_{st}= \sum_k \frac{\sigma_k\langle|\Delta_k|^2\rangle}{16 T^2 \rho_k}.
  \end{align}
  One can see that in contrast to the single band case\cite{Thompson} these
  contributions are not equal if the coupling constants are not degenerate 
  $\lambda_{11}\neq \lambda_{22}$ .
  From Eqs.(\ref{Eq:OrderParameterAmplitude}) and (\ref{Eq:kappaSingleBand}) 
  we obtain the conductivity slope at $T=T_c$
  \begin{equation} \label{Eq:S(Tc)}
  S = S_c \frac{\sum_{k} \nu_ka_k^2 D_k}{2\sum_{k} \nu_ka_k^4} 
  \sum_k \frac{ \sigma_ka_k^2}{\sigma_n} \left( \frac{1}{D_k} + 
  \frac{{\rm Tr}\hat A}{\sum_j A_{jj} D_j} \right) ,
  \end{equation}
  where $S_c= \pi^4/(14\zeta(3)\beta_L) \approx 5$ is the universal 
  value of  $S(T=T_c)$ in the single-component case.
  Let us consider a two-band system and assume that
  $\lambda_{11}>\lambda_{22}$
  and $\lambda_{12}\ll \lambda_{11}-\lambda_{12}$, which qualitatively
  corresponds to the pairing in MgB$_2$. Then the limiting cases of 
  Eq. (\ref{Eq:S(Tc)}) are as follows 
  \begin{align} \label{Eq:SD1ggD2}
  & S = \left (1+\frac{A_{22}}{2A_{11}}\right)S_c, \;\;\; 
  {\rm for} \;\; 
  D_1\gg D_2 ,  \\ \label{Eq:SD2ggD1}
  & S = \frac{\lambda_{21}^2 S_c}{2(\lambda_{11}-\lambda_{22})^2 },
  \;\;\; {\rm for}\;\; D_2\gg D_1
  \end{align}        
  These expressions are in good agreement with the behaviour of the curves
  $S(T)$ for MgB$_2$. As shown in Fig.(\ref{Fig:Conductivity})a, in the limiting case 
  $D_1\gg D_2$ (the magenta uppermost curve) the value of $S(T_c)$ is a bit 
  larger that for the single-band case, exactly as 
  described by the Eq.(\ref{Eq:SD1ggD2}) % if $\lambda_{22}<\lambda_{11}$ 
  because in this case % $A_{22}/A_{11} \approx 2\lambda_{12}\lambda_{21}/[\lambda_{22}(\lambda_{11}-\lambda_{22}] $.
  $A_{22}/A_{11} \approx (\lambda_{11}-\lambda_{22})^2/(\lambda_{12}\lambda_{21}) $.
  In the opposite case $D_2\gg D_1$ shown in Fig.(\ref{Fig:Conductivity})b 
  (magenta lowermost curve) the value $S(T_c)\ll S_c$ as given by the
  Eq.(\ref{Eq:SD2ggD1}).         
  Quite amazingly the deviations of $S(T)$ from the single-band case 
  are significant even if one of the diffusivities dominates which means
  that in the normal state the current flows mostly in one of the bands. 
  At the same time the superconducting corrections $\sigma_{st}$ and $\sigma_{fl}$ 
  are strongly renormalized by multiband effects even in the limiting cases of strong disparity between the diffusivities.  
  
  \subsubsection { Limiting values of $S$: the case of decoupled bands}
  \label{SubSubSec:LimitS}
  To understand the qualitative features of the flux-flow at high fields it 
  is instructive to consider the 
  case of superconductor with two decoupled bands characterized by different critical 
  temperatures $T_{c1,2}$. In this case superconductivity at high fields survives 
  only in one of the bands which has the 
  highest critical field  $H_{c2} = \max H_{c2}^{(k)} $. 
  Correspondingly the resistivity slope calculated for this particular band 
  coincides with the universal single-band result 
  \cite{Thompson}. However even in this case the overall $S$ is still modified 
  by multiband effects. Indeed its definition (\ref{Eq:SDefinition}) contains 
  the total normal state conductivity 
  determined by the contribution of all bands, including non-superconducting ones.
  
  Let us consider the analytically tractable low-temperature limit when 
  the single-band critical field is given by $H_{c2}^{(k)}\propto  T_{ck}/( e D_k)$ \cite{SingleBandHc2Werthamer,MakiHc2,deGennesHc2}.   
  In this case we obtain $ S=S_0 \sigma_k/(\sigma_1+\sigma_2)$,  where $k$ is the component 
  with larger critical field and $S_0=2/\beta_L\approx 1.72$ is the universal low-temperature limit 
  $S(T=0)$ in the single-component case. It is instructive to consider asymptotic behaviour of $S$ as a function of the diffusivity ratio 
  $d= D_2/D_1$. When $d<T_{c2}/T_{c1}$ we have $S =S_0 \nu_2 d/(\nu_1+ \nu_2 d)$ and $S=S_0\nu_1 /(\nu_1+ \nu_2 d)$ in the 
  opposite case when $d> T_{c2}/T_{c1}$. For non-interacting bands the transition between these regimes is abrupt resulting in the 
  jump on $S(d)$ dependence at $d=T_{c2}/T_{c1}$. The maximal value of $S$ which can be obtained does not exceed $S_0$. 
     
  The origin of a non-monotonic $S(d)$ dependence is determined by the behaviour of $H_{c2}$
  in multiband systems. At small $d$ the critical field is determined by the second band which has the smallest diffusivity 
  $H_{c2}=H^{(2)}_{c2}$. Then with increasing $d$ the superconductivity changes the host band 
  so that $H_{c2}=H^{(1)}_{c2}$. This transition is an abrupt one for non-interacting bands but a finite interaction makes 
  it the gradual one washing out the cusp singularity. However coupling does not eliminate the non-monotonicity 
  and the asymptotic behaviour of $S(d)$ remains the same as shown in 
  Fig.(\ref{Fig:ConductivityLowHighFields}).
        
 \section{Small magnetic fields $B\ll H_{c2}$ and low temperatures $T\ll T_c$. }

 \label{Sec:SmallFields}
 
 \subsection{General formalism} 

 In dilute vortex configurations at temperatures much below $T_c$ 
 the sizeable quasiparticle density exists only inside vortex cores where the 
 superconducting  order parameter is suppressed.   
 In this regime the deviations from equilibrium in each band are localized in vortex cores 
 and  are significant only at energies much smaller than the bulk energy gaps. 
 Following Kopnin-Gor'kov theory\cite{GorkovKopnin1}, spectral functions 
 $\hat g^{R,A}_k$ can be 
 parametrized at $\varepsilon =0$ as follows
 \begin{eqnarray}\label{Eq:SpectralGRA}
 & \hat g^R_k & = \tau_3\cos\theta_k + \tau_2\sin\theta_k, \\
 \nonumber
 & \hat g^A_k & = -\tau_3\cos\theta_k + \tau_2  \sin\theta_k .
 \end{eqnarray}
 Here we assume that the order parameter vortex phase is removed by gauge transformation. 
  The distribution function can be written in the form 
 \begin{equation} \label{Eq:DFansatz}
 f^{(k)}_T = \tilde{f}^{(k)}_T v_L
 \partial_\varepsilon f_0 \sin\varphi , 
 \end{equation}  
 where $\varphi$ is a polar angle  with respect to the vortex center. 
 The amplitude $\tilde{f}^{(k)}_T$ is a function
 of the radial coordinate determined by the following kinetic equation
 \begin{equation} \label{Eq:KineticFTNew}
 \left( \frac{d^2}{dr^2}+ \frac{1}{r} \frac{d}{dr}  -\frac{1}{r^2}\right)
 \tilde{f}^{(k)}_{T}  = 
 \frac{\Delta_k \sin\theta_k}{D_k} \left(2\tilde{f}^{(k)}_{T} - \frac{1}{r}
 \right) 
 \end{equation} 
 with boundary conditions $\tilde{f}^{(k)}_{T} (r=0,\infty)=0$.  
 The detailed derivation of Eq.(\ref{Eq:KineticFTNew}) is given in
 the Appendix(\ref{App:KinDerivation}).
 
 The viscous friction force acting on individual 
 moving vortices can be written as ${\bm F}_{env} = -\eta {\bm v}_L$. 
 The viscosity coefficient $\eta$ can be calculated    
substituting spectral functions (\ref{Eq:SpectralGRA}) and the 
 distribution function (\ref{Eq:DFansatz})  into the expansion (\ref{Eq:GnstExp})
 and the general expression for the force (\ref{Eq:FenvPartial}). In this way we obtain 
 \begin{align} \label{Eq:Visc}
 &\eta= \sum_k \pi\hbar\nu_k (\alpha_k + \gamma_k), \\ \nonumber
 & \alpha_k = \int_0^\infty dr r \frac{\partial \Delta_k}{\partial r}
 \frac{\partial\sin\theta_k}{\partial r}, \\ \nonumber
 & \gamma_k = \int_0^\infty dr \Delta_k \sin\theta_k \left( \frac{1}{r}-
 2\tilde{f}^{(k)}_{T} \right).
 \end{align}
 
 To calculate the gap profiles and spectral functions we use a stationary 
 self-consistency  equation written in the form  
 \begin{equation}\label{Eq:SelfConsistency2B-2}
 \Delta_i ({\bf r}) =\sum_{k=1}^N    \lambda_{ik} \left [ \Delta_k G_0 + 2\pi T 
 \sum_{n=0}^{\infty}  \left( \sin \theta^M_j - \frac{\Delta_j}{\omega_n} \right) \right] ,
 \end{equation}
 where 
 $G_0 = ( {\rm Tr} \hat\Lambda - \sqrt{ {\rm Tr} \hat\Lambda^2 - 
 4 {\rm Det} \hat\Lambda } )/(2{\rm Det} \hat\Lambda) - \ln(t)$ 
 and 
 $\hat\Lambda = \lambda_{ij}$ is the coupling matrix. 
 In Eq.(\ref{Eq:SelfConsistency2B-2}) the summation runs over Matsubara frequencies 
 $\omega_n =(2n+1)\pi T$. The angle $\theta^M_k$  parametrizes 
 imaginary-frequency Green's functions similar to Eqs.(\ref{Eq:SpectralGRA}).  
 It is determined by the Usadel equation 
 \begin{equation}\label{Eq:teta-Usadel}
 \frac{1}{r} \frac{d}{dr} \left( r \frac{d\theta^M_k}{dr} \right)  - 
 \frac{\sin(2\theta^M_k)}{2r^2} +
 \frac{2\Delta_k}{D_k}\cos\theta^M_k-\frac{2\omega}{D_k}\sin \theta^M_k=0,
 \end{equation}
 supplemented by the boundary conditions 
 \begin{align}\label{Bc-Usadel}
 &\theta^M_k(r=0)=0,\\
 &\theta^M_k(r=\infty)= \sin^{-1} \left[ \Delta_k/
 \left(\Delta_k^2+\omega^2\right)^{1/2} \right] . \nonumber
 \end{align}
 One should put $\omega=\omega_n$ to obtain solutions at the specific Matsubara frequency. 
 The angle $\theta_k$ parametrizing  zero-energy spectral functions (\ref{Eq:SpectralGRA}) 
 is given by the same Eqs.(\ref{Eq:teta-Usadel},\ref{Bc-Usadel}) with $\omega=0$. 
  
 In general the flux-flow conductivity can be expressed through the vortex
 viscosity (\ref{Eq:Visc}) as follows \cite{GorkovKopnin1}
 \begin{equation}\label{Eq:FluxFlowCond}
 \sigma_f= \eta /(B\phi_0),
 \end{equation}
 where $B$ is the average magnetic induction and $\phi_0$ is a magnetic flux quantum.  
 Introducing normal-state Drude conductivity  $\sigma_n=\sum_k\sigma_k$, we rewrite 
 flux-flow conductivity (\ref{Eq:FluxFlowCond}) in the form
 \begin{align} \label{Eq:FluxFlowCondVisc}
 & \sigma_f= \beta \sigma_n H_{c2}/B, \\ \label{Eq:beta}
 & \beta= \frac{1}{2eH_{c2}}\frac{\sum_k \nu_k (\alpha_k + \gamma_k)}{\sum_k \nu_kD_k}.
 \end{align}
 In Sec.\ref{Sec:ExamplesComparison} we analyse parameter $\beta$ for several 
 known multiband compounds superconducting compounds.
   
 %%%%%%%%%%%%%%%%%%%%%%%%%%%%%%%%%%%%%%%%%%%%%%%%%%%%%%%%%%%%%%%%%%%%%%%%%
 \subsection{Exact value of $\beta$ in a single-component case}
 \label{SubSec:ExactBeta} 
\begin{figure}[t!]
 \includegraphics[width=0.85\linewidth]{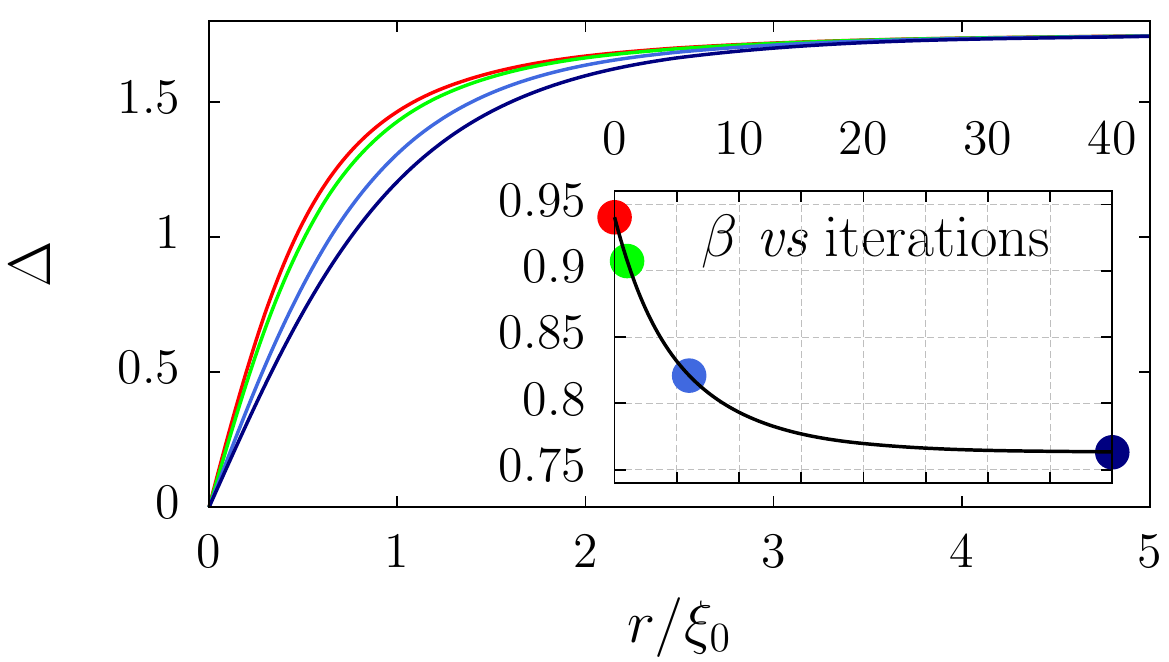}
 \caption{\label{f1} 
 (Color online) { Single vortex solution of one-band self-consistency equation
 solved by iterations. The initial distribution (red) given by the Clem ansatz
 and first iteration (green) used in \cite{GorkovKopnin1} are shown
 compared to 6th (light blue) and 40th (dark blue) iterations. The flux-flow
 conductivity slope $\beta$ is depicted in inset as function of iteration number.
 Values of $\beta$ which corresponds to the gap distributions shown are indicated
 in the inset by dots with analogous colour. Here, gap order parameter is scaled
 by $\Delta_\mathrm{bulk}/T_\mathrm{c}$ and distance by
 $\xi_0=(\hbar D/T_\mathrm{c})^{1/2}$.}}
 \end{figure}

 As can be seen from the Eqs.(\ref{Eq:beta}) in the single-component case the value of $\beta$ is fixed. 
 Previously the value of $\beta\approx 0.9$ has been reported\cite{GorkovKopnin1} calculated based on the 
 approximate distribution of the order parameter near a vortex \cite{watts-tobin}. The vortex structure 
 was obtained by a single iteration of the self-consistency Eq.(\ref{Eq:SelfConsistency2B-2}) with $N=1$ 
 starting from the Clem-Hao initial guess\cite{clem}. By performing sufficient iterations of self-consistency equation, 
 one can ascertain that within weak-coupling limit the fully self-consistent vortex structure yields an exact 
 value of $\beta=0.76$. Fig. \ref{f1} demonstrates disparity between initial gap distribution, first iteration 
 and exact gap function together with corresponding values of $\beta$. 
 %%%%%%%%%%%%%%%%%%%%%%%%%%%%%%%%%%%%%%%%%%%%%%%%%%%%
 %%%%%%%%%%%%%%%%%%%%%%%%%%%%%%%%%%%%%%%%%%%%%%%%%%%%   
 
\subsection{Limiting values of $\beta$: the case of decoupled bands}
 \label{SubSec:DecoupledBandsBeta}
 In multiband superconductors, the flux-flow conductivity behaviour is more involved
 so that the coefficient $\beta$ can change a lot depending on the ratio of
 the diffusion constants and pairing potentials in different bands. 
 Below we investigate the maximal accessible values and the asymptotic behaviour of $\beta$ 
 in superconductors with decoupled bands characterized by different critical temperatures 
 $T_{ck}$. In this case one can adopt single-band results discussed in the previous section
 \ref{SubSec:ExactBeta} to obtain 
 \begin{equation}
 \beta=\beta_0 \min \left( \frac{D_k}{T_{ck}} \right) \frac{\nu_1T_{c1} + \nu_2T_{c2} }{\nu_1D_{1} + \nu_2D_{2}},
 \end{equation}  
  where $\beta_0\approx 0.76$ is the single-band value. Here we have used the same single-band expression for 
  $H_{c2}=\max H_{c2}^{(k)}$ as in the section (\ref{SubSubSec:LimitS}).
  
  It is instructive to consider asymptotic behaviour of $\beta$ as a function of the diffusivity ratio 
  $d= D_2/D_1$. When $d<T_{c2}/T_{c1}$ we have 
  $\beta=\beta_0 d(\nu_2 + \nu_1T_{c1}/T_{c2} )/(\nu_1+ \nu_2 d)$ and 
  $\beta=\beta_0 (\nu_1 + \nu_2T_{c2}/T_{c1} )/(\nu_1+ \nu_2 d)$ in the 
  opposite case when $d>T_{c2}/T_{c1}$. For non-interacting bands the transition between these regimes is abrupt resulting in the 
  sharp maximum $\beta = \beta_0$ with a cusp at $d= T_{c2}/T_{c1}$. The transition results from switching of the superconductivity 
  at $H_{c2}$ between different bands. If there is a  finite interband coupling, the cusp in the behaviour of $\beta$ changes 
  to smooth maximum but the maximal value cannot be remarkably enhanced. As a result, parameter $\beta$ in two-band scenario 
  appears to be always limited by its single-band value.

 %%%%%%%%%%%%%%%%%%%%%%%%%%%%%%%%%%%%%%%%%%%%%%%%%%%% 
 %%%%%%%%%%%%%%%%%%%%%%%%%%%%%%%%%%%%%%%%%%%%%%%%%%%%
 
 %%%%%%%%%%%%%%%%%%%%%%%%%%%%%%%%%%%%%%%%%%%%%%%%%%%%%%%%%%%%%%%%%%%%%%%%%  
 \section{Examples and comparison with experiments }
 \label{Sec:ExamplesComparison}
  
  Having in hand general results we can calculate the flux-flow resistivity in
  particular multiband superconducting compounds. For that we choose MgB$_2$
  and {V$_3$Si} which have been described by the two-band weak coupling
  models\cite{KoganMgB2,GurevichMgB2,KoshelevGolubovMgB2,V3Si}. 
  Moreover these compounds can have rather large impurity scattering rate to 
  fit the dirty limit conditions\cite{GurevichMgB2,KoshelevGolubovMgB2}. 
  
  %%%%%%%%%%%%%%%%%%%%%%%%%%%%%
  Basically the only input parameters needed to calculate the flux-flow resistivity  are the pairing constants which we choose as follows  
  {\bf (i)} MgB$_2$ with $\lambda_{11} =
  0.101$, $\lambda_{22} = 0.045$, $\lambda_{12} = 0.034$, $\lambda_{21} = 0.026$
  \cite{KoganMgB2} and {\bf (ii)} {V$_3$Si with $\lambda_{11} = 0.26$}, 
  $\lambda_{22} = 0.205$, $\lambda_{12}=\lambda_{21} = 0.0088$ \cite{V3Si}.
  Note that the parameters of {V$_3$Si} correspond to the case of weakly interacting 
  superconducting bands since the interband pairing is much weaker than the intraband one
  $\lambda_{12}\ll \lambda_{11},\lambda_{22}$.
  In that sense it is drastically different from the model of MgB$_2$ where 
  $\lambda_{12}$ has the same order of magnitude as $\lambda_{11}$ and $\lambda_{22}$.
    
\begin{figure}[h!] \includegraphics[width=1.0\linewidth]{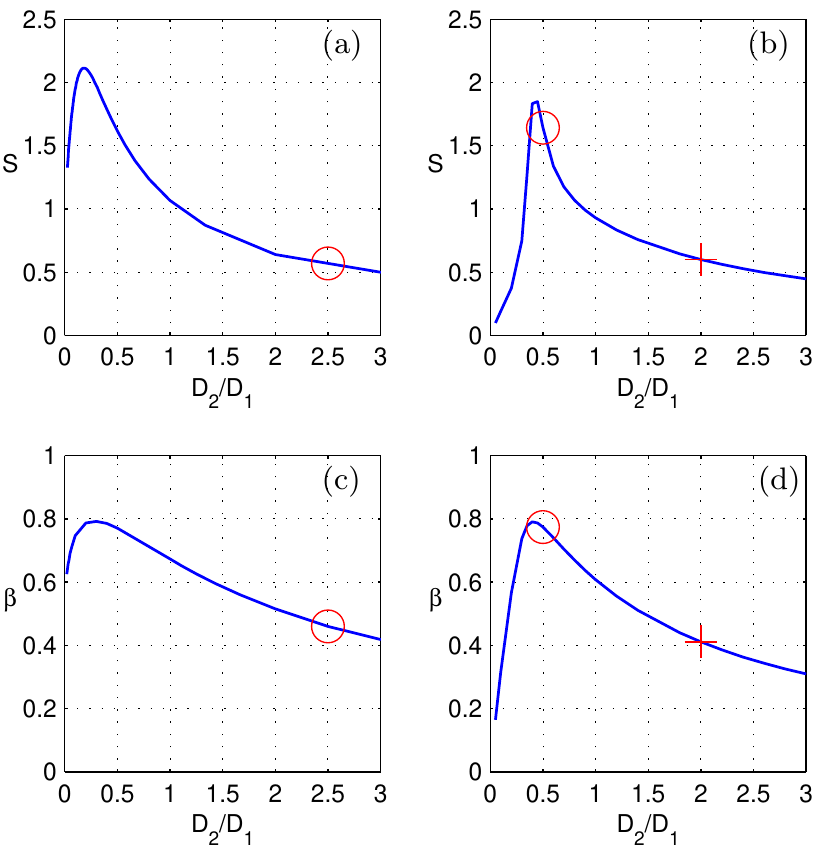} 
  \caption{\label{Fig:ConductivityLowHighFields} (Color online)
  The flux flow resistivity slope $S$ at high fields (\ref{Eq:SDefinition}) and
  the inverse slope $\beta$ at low fields (\ref{Eq:BetaDefinition})
   in a two-band superconductor
  as functions of the diffusivity ratio  $D_2/D_1$. 
  The temperature is $T=0.05 T_c$.
  The panels correspond to the models of (a,c) MgB$_2$ \cite{KoganMgB2} and 
  (b,d)  V$_3$Si \cite{V3Si} with the pairing constants mentioned in the text.
  Open circles in (a,c) mark the parameters used to fit the experimental curve for 
  MgB$_2$ in the Fig.(\ref{Fig:CondFit})a.
  Open circles and crosses in (b,d) show the parameters used to extrapolate magnetoresistance curves for 
  V$_3$Si in the Fig.(\ref{Fig:CondFit})b below.  }
 \end{figure}

  For such parameters we apply the results of sections (\ref{Sec:SHc2}) and (\ref{Sec:SmallFields}) to find the dependencies 
  {$S(d)$} and $\beta(d)$ where $d=D_2/D_1$ is the ratio of diffusivities in the two bands. The results are shown in 
  Fig.(\ref{Fig:ConductivityLowHighFields}). On can see that the dependencies are qualitatively similar for the two sets of pairing constants. 
  The non-monotonicity and asymptotic behaviour of both $S$ and $\beta$ were explained in sections (\ref{SubSubSec:LimitS}) and (\ref{SubSec:DecoupledBandsBeta})
  using a model of non-interacting bands.
   As was discussed above the origin of a non-monotonic behaviour is determined by the multiband effects 
  in the near-$H_{c2}$ physics. In that regime increasing the ratio $D_2/D_1$ one makes the superconductivity to change host band. 
  This affects directly the resistive states at high fields (i.e. the $S$ parameter) but also indirectly 
  the low-field parameter $\beta$ (\ref{Eq:BetaDefinition}) because there 
  the magnetic field dependence is normalized by $H_{c2}$.  
  
 %%%%%%%%%%%%%%%%%%%%%%%%%%%%%%%%%%%%%%%%%%%%%%%%%%%%
  %%%%%%%%%%%%%%%%%%%%%%%%%%%%%%%%%%%%%%%%%%%%%%%%%%%% 
  
\begin{figure}[h!] \includegraphics[width=1.0\linewidth]{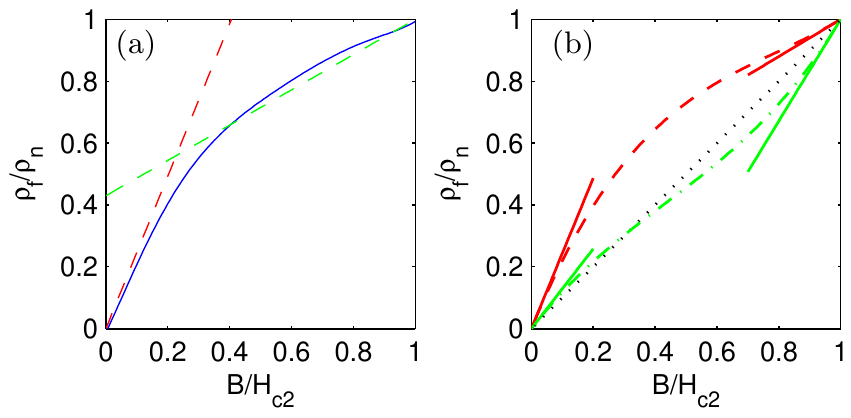} 
  \caption{\label{Fig:CondFit} (Color online)
 (a) Solid line: the approximate flux flow resistivity as a function of magnetic induction $B$ adopted from the experimental curve for MgB$_2$\cite{FluxFlowExperimentsMgB2}. Dashed lines show theoretical slopes
 $\beta=0.406$ and $S=0.57$  corresponding to $D_2/D_1= 2.5$ for the same  model as in  Fig.(\ref{Fig:ConductivityLowHighFields})a,c, where 
 these slopes are shown by circles. 
 (b) The slopes of resistivity corresponding to 
   $D_2/D_2 =0.5$ (dash-dotted line), and $D_2/D_1 =2$ (dashed line) for the same model of V$_3$Si is in the Fig.(\ref{Fig:ConductivityLowHighFields})b,d. The solid curves are cubic interpolation between the low and high field regimes. 
   The temperature is $T=0.05 T_c$ in both panels. }
 \end{figure}

  With the help of calculated parameters $S$  and $\beta$ we can reconstruct by extrapolation the flux-flow resistivity  
  curve for the entire range of magnetic field and compare it with the experimental results. 
  In Fig.(\ref{Fig:CondFit})a we show by dashed lines the slopes that give the best fit 
  of the approximated flux-flow resistivity curve for MgB$2$ at low temperatures $T\ll T_c$, adopted from  
  Ref. (\onlinecite{FluxFlowExperimentsMgB2}). The slopes were calculated using the two-band model for MgB$_2$ described above.
   The fitting parameter was the ratio of diffusivities chosen to be $D_2/D_1=2.5$ 
  marked in the Fig.(\ref{Fig:ConductivityLowHighFields})a,c by open circles.
   
  To understand the possible variations in the shape of the curve $\rho_f(B)$ we consider the model corresponding to V$_3$Si
   and consider two characteristic values of $D_2/D_1=2$ and $0.5$. For such parameters the values of $S$ and $\beta$ are shown by 
   open circles and crosses in the Fig.(\ref{Fig:ConductivityLowHighFields})b,d. One can see that one of these points is in the  
   regime qualitatively similar to the one considered above for MgB$_2$. Indeed the cubic extrapolation of the $\rho_f(B)$ dependence shown 
   by red dashed curve is qualitatively similar to the approximated experimental curve for MgB$_2$, see 
   Fig.(\ref{Fig:CondFit}). On the other hand, the point $D_2/D_1=0.5$ belongs to the region where $S>1$,  which results in a different behaviour shown in Fig.(\ref{Fig:CondFit})b with green color. Experimental data for V$_3$Si \cite{V3Si_2} demonstrates flux-flow resistivity curve between two cases considered, however more measurements are needed to cover whole range of magnetic fields. Note that green curve in Fig. (\ref{Fig:CondFit})b 
  is quite close to the usual BS linear dependence (black dotted curve) although deflects slightly changing its shape from the concave
   at small fields to the convex one at large fields. Slightly varying ratio $D_2/D_1$ around $0.5$, one can achieve better approach to BS line. At the same time since $\beta$ does not exceed much its single-band limit value $\beta_0=0.76$ it is impossible to get a convex curve already at small fields since that would require $\beta>1$ which we did not obtain for the models considered in the present work.

 %%%%%%%%%%%%%%%%%%%%%%%%%%%%%%%%%%%%%%%%%%%%%%%%%%%%
 %%%%%%%%%%%%%%%%%%%%%%%%%%%%%%%%%%%%%%%%%%%%%%%%%%%%
  
 \section{Summary} \label{Sec:Conclusion}
  To summarize we have developed theoretical framework to study non-equilibrium processes in multiband superconductors
  and applied it to calculate flux-flow resistivity of such systems in the dirty limit with high concentration 
  of non-magnetic impurities. We have considered both the regions of high and and low magnetic fields.
  To calculate the conductivity in the former case we have derived the solution charactering moving vortex 
  lattices which reveals the effect of splitting into sublattices of fractional vortices. The maximal value of the flux-flow resistivity 
  slope $S$ at high fields is shown to be close to the universal single-band limit. At the same time the minimal value can be arbitrary small
  proportional to the disparity of diffusivities in different bands %$S\propto \min(D_1, D_2)/D_1D_2$.  
  $S\propto \min(D_{1,2})/\max(D_{1,2})$.
  
  We calculated the parameter $\beta$ which is inverse 
  slope of flux-flow magneto-resistance curve 
  at low magnetic field. For different models of multiband superconductors
 we have found that the maximal possible value of $\beta$ is close to the the universal single-band constant $\beta_0$.
 The approximate value of $\beta_0$ was found by Kopnin and Gor'kov\cite{GorkovKopnin1}. The exact value calculated in the present work is 
 $\beta_0=0.76$. For large disparity of diffusivities it has the asymptotic behaviour % $\beta\propto \min(D_1, D_2)/D_1D_2 $
 $\beta\propto\min(D_{1,2})/\max(D_{1,2})$ which is similar to that of parameter $S$. 
 
 We demonstrated that multiband superconductors exhibit unconventional generic regime which is characterized by small
 values of parameters $\beta$ and $S$ and corresponds to the concave flux-flow magnetoresistance curves $\rho_f(B)$. Several recent experiments\cite{FluxFlowExperimentsFeAs1, FluxFlowExperimentsFeAs2, FluxFlowExperimentsFeAs3,FluxFlowExperimentsFeAs4,FluxFlowExperimentsFeAs5, FluxFlowExperimentsMgB2} confirm that behaviour. For MgB$_2$ we have obtained a quantitative agreement with experimental results \cite{FluxFlowExperimentsMgB2}
 choosing the ratio of diffusivities in two bands $D_2/D_1=2.5$. At the same time we have shown that by varying the  parameter 
 $D_2/D_1$ it is possible to obtain regimes when the curve $\rho_f(B)$ is quite close to the single-band Bardeen-Stephen law. 
 Therefore, the suggested theory naturally explains quite diverse experimental data on the flux-flow resistivity in different mutliband superconducting compounds.

 \section{Acknowlegements} 
 The work was supported by the Estonian Ministry of Education and Research (grant PUTJD141),
 Goran Gustafsson Foundation  and by the Swedish
Research Council grant 642-2013-7837.

 \appendix

 \section{Derivation of kinetic equations and 
 forces acting on the moving vortex line} \label{App:KinDerivation}

 \subsection{General formalism}
 The quasiclassical GF matrix in band $k$ is defined as
 \begin{equation}
 \check{g}_k = \left(%
 \begin{array}{cc}
 \hat g^R_k &  \hat g^K_k \\
 0 &  \hat g^A_k \\
 \end{array}\label{eq:GF0App}
 \right)\; ,
 \end{equation}
 where $g^K_k$ is the (2$\times$2 matrix) Keldysh component 
 and $\hat g^{R(A)}_k$ is the retarded (advanced) GF. In a diffusive
superconducting wire with band diffusion constants $D_k$ the matrix $\check g_k$
 obeys the Usadel equation 
 \begin{equation}\label{Eq:KeldyshUsadelApp}
 \{\tau_3\partial_t, \check g_k \}_t = D_k\hat\partial_{\bf r}  ( \check g_k
\circ \hat\partial_{\bf r} \check g_k) 
 + [\hat H_k , \check g_k ]_t -i [\check \Sigma^{ph}_k , \check g_k ]_t,
  \end{equation} 
 where 
 $\hat H_k ({\bm r},t) = i \hat \Delta_k $ and $\hat\Delta_k (t)= |\Delta_k|
\sigma_3\tau_3\tau_1e^{-i\varphi_k \tau_3}$ is the gap operator and $\varphi_k$
is 
 the gap phase. It is convenient to remove the spin dependence of gap by
  transformation $\check{g}=\check{U} \check{g}^{new} \check{U}^+$
  where
  \begin{equation}\label{Eq:Transformation}
  \check{U}=\exp  \left[ i\pi
  (\sigma_3\tau_3-\sigma_0\tau_3-\sigma_3\tau_0)/4\right ],
  \end{equation}
  which leads to
  \begin{equation}\label{Eq:DeltaNew}
  \check{\Delta}^{new}_k= \check{U}^+\check{\Delta}_k \check{U}= 
  i |\Delta_k|\tau_2 e^{-i\varphi_k \tau_3},
  \end{equation}
 so that $\hat H_k ({\bm r},t) = i \hat \Delta_k^{new}$. Note that we use from
 the beginning the temporal gauge where the scalar potential 
 is zero $\Phi=0$ with and additional constraint that
 in equilibrium the vector potential is time-independent and satisfies
 $\nabla\cdot\bm A=0$. Throughout the derivation we assume $k_B=\hbar=c=1$.

 The covariant differential superoperator in Eq. (\ref{Eq:KeldyshUsadelApp}) 
 is given by 
  $$
  \hat \partial_{\bf r} \hat g_k= \nabla \hat g_k -ie\left[\tau_3{\bm A}, 
  \hat g_k \right]_t.
  $$
  Here the commutator operator is defined as
  $[X, g]_t= X(t_1) g(t_1,t_2)- g(t_1,t_2) X(t_2)$, similarly for 
  anticommutator $\{X,g\}_t$.
   We also introduce the symbolic product operator 
   $ (A\circ B) (t_1,t_2) = \int dt A(t_1,t) B(t,t_2) $. Equation (\ref{Eq:KeldyshUsadelApp}) 
   is complemented by the normalization condition
   $(\check g_k\circ \check g_k) (t_1,t_2)= \check \delta (t_1 - t_2)$
   which allows to introduce parametrization of the Keldysh component in terms of the 
   distribution function
   \begin{eqnarray}
   \hat g^K_k = \hat g^R_k\circ \hat f^{(k)}- \hat f^{(k)}\circ \hat g^A_k, \\ 
   \label{Eq:ParametrizationApp}
   \hat f^{(k)}= f_L^{(k)}\tau_0 +f_T^{(k)}\tau_3 .  \label{Eq:DistrFunApp}
   \end{eqnarray}
   
 Here we will neglect the electron-phonon relaxation given by the last term in
 the Eq.(\ref{Eq:KeldyshUsadelApp}). Such approximation is valid provided the temperature is not too close to $T_c$. 
 In this case the components of the
 Keldysh-Usadel Eq. (\ref{Eq:KeldyshUsadelApp}) read as
 \begin{align}\label{Eq:KeldyshUsadelKeldyshComponent}
 & \{\tau_3\partial_t, \hat g^{R(A)}_k \}_t = 
 D_k\hat\partial_{\bf r} ( \hat g^{R(A)}_k \circ \hat\partial_{\bf r} \hat
g^{R(A)}_k ) + [\hat H_k , \hat g^{R(A)}_k ]_t, \nonumber \\
 & \{\tau_3\partial_t,  \hat g^K_k \}_t = D_k\hat\partial_{\bf r} ( \hat g^R_k
\circ \hat\partial_{\bf r} \hat g^K_k +
 \hat g^K_k \circ \hat\partial_{\bf r} \hat g^A_k  ) + [\hat H_k , \hat g^K_k
]_t.
 \end{align}   
  To obtain kinetic equation we substitute parametrization
(\ref{Eq:Parametrization}) to write
 \begin{eqnarray}
 \hat\partial_{\bf r} ( \check g_k \circ \hat\partial_{\bf r} \check g_k)^K = 
 \hat\partial_{\bf r} (\hat\partial_{\bf r} \hat f^{(k)} - \hat g^R_k\circ
\hat\partial_{\bf r}\hat f^{(k)}\circ \hat g^A_k )+ \\ \nonumber
 \hat g^R_k\circ\hat\partial_{\bf r} \hat g^R_k\circ \hat\partial_{\bf r} \hat
f^{(k)} - 
 \hat\partial_{\bf r} \hat f^{(k)}\circ \hat g^A_k\circ\hat\partial_{\bf r} \hat
g^A_k +  \\ \nonumber
 \hat\partial_{\bf r} (\hat g^R_k\circ\hat\partial_{\bf r} \hat g^R_k )\circ
\hat f^{(k)} - \hat f^{(k)} \circ \hat\partial_{\bf r}
 (\hat g^A_k\circ\hat\partial_{\bf r} \hat g^A_k ) .
 \end{eqnarray}  
 To derive this expression we used the associative property of differential
superoperator
 $
 \hat\partial_{\bf r}(g_1\circ g_2)= \hat\partial_{\bf r}g_1\circ g_2 + g_1\circ
\hat\partial_{\bf r} g_2
 $.  To get rid of the last two terms we subtract the spectral components of the
Eq.(\ref{Eq:KeldyshUsadelApp}) to obtain finally
 the equation
  \begin{eqnarray} \label{Eq:KineticEqGen}
  & \hat g^R_k \circ(\tau_3 \partial_{t^\prime} \hat f^{(k)}+\partial_{t_2}\hat
f^{(k)}\tau_3)-\nonumber\\
  &(\tau_3\partial_{t_1}\hat f^{(k)}+\partial_{t^\prime}\hat
f^{(k)}\tau_3)\circ\hat g^A_k=\nonumber \\  \nonumber
  & D_k\hat\partial_{\bf r} (\hat\partial_{\bf r} \hat f^{(k)} - \hat g^R_k\circ
\hat\partial_{\bf r}\hat f^{(k)}\circ \hat g^A_k )+\\ \nonumber
  & D_k ( \hat g^R_k\circ\hat\partial_{\bf r} \hat g^R_k\circ \hat\partial_{\bf
r} \hat f^{(k)} - 
   \hat\partial_{\bf r} \hat f^{(k)}\circ \hat g^A_k\circ\hat\partial_{\bf r}
\hat g^A_k )+\nonumber\\
   & \hat g^R_k \circ [\hat H_k, \hat f^{(k)}]_t - [\hat H_k, \hat f^{(k)}]_t
\circ\hat g^A_k,
  \end{eqnarray}
where $t^\prime$ is integration variable.

  To proceed we introduce the mixed representation in the time-energy domain as
follows
  $g (t_1,t_2) = \int_{-\infty}^{\infty} g(\varepsilon, t) e^{-i\varepsilon
(t_1-t_2) } 
  \frac{d\varepsilon}{2\pi} $, 
  where $t=(t_1+t_2)/2$. 
 By keeping the first order terms in frequency, we get for Fourier
transformations
  \begin{align}
  [\hat H , \hat g ]_t = [\hat H , \hat g ] - \frac{i}{2} \{ \partial_t
  \hat H, \partial_\varepsilon\hat g\}, \\
  [\bm A \tau_3, \hat g ]_t = \bm A[ \tau_3, \hat g ] - 
  \frac{i}{2} \partial_t \bm A \{ \tau_3, \partial_\varepsilon\hat g \}, \\
 \hat\partial_{\bf r} {\hat f}^{(k)}  = \nabla \hat{f}^{(k)} + e{\bm E}
\partial_\varepsilon f_0\tau_3,
  \end{align}
 where ${\bm E}=-\partial_t{\bm A}$ is electric field in temporal gauge and
$f_0=\tanh\varepsilon/(2T)$ is equilibrium distribution.
To the first order in frequency and deviation from equilibrium we also have
  \begin{eqnarray} \nonumber
  \hat\partial_{\bf r} (\hat\partial_{\bf r} \hat f^{(k)} - \hat g^R_k\circ
\hat\partial_{\bf r}\hat f^{(k)}\circ \hat g^A_k )  = \\ \nonumber
  \nabla(\nabla f^{(k)} - \hat g^R_k \nabla f^{(k)} \hat g^A_k) + e
\partial_\varepsilon f_0 \nabla \cdot ({\bm E} (\tau_3- \hat g^R_k \tau_3\hat
g^A_k)) \\ \nonumber
+ ie [{\bm A}\tau_3, \hat g^R_k \nabla \hat f^{(k)} \hat g^A_k ] + i e^2
\partial_\varepsilon f_0 ({\bm A}\cdot {\bm E})
  [\tau_3, \hat g^R_k \tau_3 \hat g^A_k ].
  \end{eqnarray}         
   The last to terms do not contribute to the kinetic equation since they are
traced out.  

  In the mixed representation the kinetic Eq.(\ref{Eq:KineticEqGen}) has the
following gauge-invariant form
  \begin{eqnarray} \label{Eq:KineticEq}
  & \hat g^R_k \tau_3 \partial_t \hat f^{(k)} - \tau_3\partial_t \hat f^{(k)}
\hat g^A_k = D_k \nabla(\nabla f^{(k)} - \hat g^R_k \nabla f^{(k)} \hat g^A_k)
+\nonumber\\
  & D_k ( \hat g^R_k\hat\partial_{\bf r} \hat g^R_k \nabla \hat f^{(k)} - 
   \nabla \hat f^{(k)} \hat g^A_k\hat\partial_{\bf r} \hat g^A_k ) +  \hat g^R_k
[\hat H_k, \hat f^{(k)}] -\nonumber\\
   & [\hat H_k, \hat f^{(k)}]\hat g^A_k  - i \partial_\varepsilon f_0 ( \hat
g^R_k \partial_t\hat H_k  - \partial_t\hat H_k \hat g^A_k ) + \\ \nonumber
  & e D_k \partial_\varepsilon f_0  \nabla \cdot \left({\bm E} (\tau_3- \hat
g^R_k \tau_3\hat g^A_k)\right) + \\ \nonumber
  & e D_k \partial_\varepsilon f_0 {\bm E}\cdot ( \hat g^R_k\hat\partial_{\bf r}
\hat g^R_k \tau_3 - \tau_3 \hat g^A_k\hat\partial_{\bf r} \hat g^A_k ),
 % -  \hat I^{ph} 
  \end{eqnarray}
   where we omit the terms which will be traced out later. The last two %three
    terms in Eq. (\ref{Eq:KineticEq}) are the sources of disequilibrium.
 Multiplying by $\tau_3$ and taking the trace we obtain
 \begin{align} \nonumber
 &   \nabla ( {\cal D}_T^{(k)} \nabla f_T^{(k)} )+  {\bm j}_e^{(k)}\cdot\nabla
f_L^{(k)} + 
 2i {\rm Tr} [ (\hat g^R_k + g^A_k) \hat \Delta_k ] f_T^{(k)} = \\
\label{Eq:KineticEqFT3App} 
 & \partial_\varepsilon f_0 {\rm Tr} [ \tau_3  \partial_t\hat \Delta_k( \hat
g^R_k + \hat g^A_k) ] -
 e \partial_\varepsilon f_0 \nabla \cdot ({\cal D}_T^{(k)} {\bm E} ), 
 \end{align} 
  where the energy dependent diffusion coefficients and the spectral charge
currents are 
 \begin{eqnarray}
 {\cal D}_T^{(k)} = D_k {\rm Tr}( \tau_0 - \tau_3\hat g^R_k \tau_3\hat g^A_k ),
\\
  {\bm j}_e^{(k)} = D_k {\rm Tr}\; [ \tau_3( \hat g^R_k\hat\partial_{\bf r} \hat
g^R_k - \hat g^A_k\hat\partial_{\bf r} \hat g^A_k)].
 \end{eqnarray}  
 Analogously taking just the trace of Eq.(\ref{Eq:KineticEq}) we obtain 
  \begin{align}\label{Eq:KineticEqFLApp} \nonumber
  &\nabla ( {\cal D}_L^{(k)} \nabla f_L^{(k)} ) + {\bm j}_e^{(k)}\cdot\nabla
f_T^{(k)} + 2i {\rm Tr} [\tau_3 (\hat g^R_k -g^A_k) \hat \Delta_k ] f_T^{(k)} =
\\  
  & - \partial_\varepsilon f_0 {\rm Tr}[ \partial_t\hat \Delta_k( \hat g^R_k -
\hat g^A_k) ] - 
  e \partial_\varepsilon f_0  \left (  {\bm j}_e^{(k)} \cdot {\bm E} \right),
   \end{align} 
  where ${\cal D}_L^{(k)} = D_k  {\rm Tr} (\tau_0 - \hat g^R_k \hat g^A_k )$.
Here we took into account that ${\rm Tr} ( \hat g^R_k \tau_3\hat g^A_k) =0$
  because of the relation $\hat g^A_k= -\tau_3 \hat g^{R+}_k \tau_3$
  and the general form of the equilibrium spectral function $\hat
g^R_k=g_3^{(k)}\tau_3 + g_2^{(k)}\tau_2 e^{-i\varphi_k\tau_3}$.            
 \subsection{The low temperature limit $T\ll T_c$. }
    
 At low temperatures the deviations from equilibrium are localized in the vortex
core and are significant only at small energies. 
 Therefore following Kopnin-Gor'kov theory we can use the spectral functions
$\hat g^{R,A}_k$ calculated at $\varepsilon =0$ when it is possible 
 to use  $\theta$ parametrization in each band
 \begin{eqnarray}\label{Eq:appendixSpectralGRA}
  \hat g^R_k = \tau_3\cos\theta_k + \tau_2 e^{-i\tau_3 \varphi_k} \sin\theta_k,
\\ \nonumber
 \hat g^A_k = -\tau_3 \hat g^{R+}_k \tau_3.
 \end{eqnarray}
 %%%%%%%%%%%%%%%%%%%%%%%%%%%%%%%%%%%%%%%%%% 
 In this case we can simplify kinetic equation with the help of the following
identities 
 ${\cal D}^{(k)}_T =4D_k$ and
   \begin{eqnarray}
   2i {\rm Tr} [ (\hat g^R_k + \hat g^A_k) \hat \Delta_k ] = -8 |\Delta_k|
\sin\theta_k, \\ 
   {\rm Tr} [ \tau_3  \partial_t\hat \Delta_k( \hat g^R_k + \hat g^A_k) ] =
4({\bm v}_L\cdot \nabla\varphi_k) |\Delta_k| \sin\theta_k,
   \end{eqnarray}
  where we took into account that for the vortex moving with constant velocity
$\partial_t\Delta_k=- {\bm v}_L\cdot\nabla\Delta_k$. 
  Hence the kinetic equation becomes
  \begin{equation} \label{Eq:KineticFT}
   D_k \nabla^2 f^{(k)}_{T} = 
  \left[ 2 f^{(k)}_T + \partial_\varepsilon f_0 ({\bm v}_L \nabla\varphi_k)
\right] 
  |\Delta_k|\sin\theta_k.
  \end{equation}
   
To calculate the force ${\bm F}_{env}$ (\ref{Eq:FenvGen}) we use the expansion
(\ref{Eq:GnstExp})    
  substituting there the spectral functions in the form
(\ref{Eq:appendixSpectralGRA}) to obtain  
 \begin{align}
 {\rm Tr} (\hat g^{nst}_k \hat \partial_{\bm r} \hat \Delta_k) = 
 -2 \partial_\varepsilon f_0 [({\bm v}_L \nabla\sin\theta_k) ] \nabla|\Delta_k|
\\ \nonumber
 -2\sin\theta_k |\Delta_k|\left[ 2f^{(k)}_T + \partial_\varepsilon f_0 ({\bm
v}_L \nabla\varphi_k) \right]\nabla\varphi_k .
 \end{align}  
For small magnetic fields $B\ll H_{c2}$ the last term in Eq.(\ref{Eq:FenvGen})
can be neglected 
 so that the force is given by 
 \begin{align} \label{} \nonumber
 & {\bm F}_{env} = -\sum_k \frac{\nu_k }{2}\int d^2 {\bm r} d\varepsilon  \Big\{
 \partial_\varepsilon f_0 \nabla|\Delta_k| 
 ({\bm v}_L \nabla\sin\theta_k) + \\ 
 & |\Delta_k|\sin\theta_k \left[ 2f^{(k)}_T + \partial_\varepsilon f_0 ({\bm
v}_L \nabla \varphi_k) \right]\nabla\varphi_k\Big\}.   
 \end{align}   
  We can simplify equations further taking into account common phase
$\varphi_{1,2}=\varphi$ so that $({\bm v}_L\nabla\varphi) =-v_L\sin\varphi /r $
and 
  $({\bm v}_L\nabla|\Delta_k|) = v_L \cos\varphi \partial_r |\Delta_k|$. By
factorizing the angular dependence of the distribution function
  $f^{(k)}_T = \tilde{f}^{(k)}_T v_L \partial_\varepsilon f_0 \sin\varphi$, the
force becomes 
  ${\bm F}_{env} = -\eta {\bm v}_L$ where the viscosity coefficient is given by
(\ref{Eq:Visc}).

 \end{document}